\documentclass[11pt]{article}
\usepackage{array, amsmath, amssymb}
\begin{document}
\begin{center}
{\Large\bf Dimensional Analysis and Physical Laws}\\
\vspace{0.5cm}
{\large\bf Shinji Tanimoto}\\
(tanimoto@cc.kochi-wu.ac.jp)\\
\vspace{0.5cm}
Department of Mathematics\\
Kochi Joshi University\\
Kochi 780-8515, 
Japan. \\
\end{center}
\begin{abstract}
Dimensional analysis provides many simple and useful tools for various situations in science. 
The objective of this paper is to investigate its relations to functions, i.e.,
the dimensions for functions that yield physical quantities and those for the arguments 
of functions. In particular, it is shown that there are three types of functions
from a viewpoint of dimensional analysis. Relationships between
Planck's constant and dimensional analysis are also considered.\\
\end{abstract}
\vspace{0.7cm}
\noindent
{\bf 1.  Introduction} \\
\\
\indent
Dimensional analysis is a very simple but powerful toolbox to infer or to verify physical formula. 
Plenty of examples of its applications can be found in books [1, 2]. 
The aim of this paper is to investigate another aspect of 
dimensional analysis, which focuses on its relations to various functions. The author has not encountered treatises 
or papers that touch upon the topic explicitly, 
although it is also helpful in order to infer or to verify physical formula. 
Actually it will be of benefit to know, for instance, that the arguments of exponential and trigonometric functions
must be dimensionless. \\
\indent
We deal with phenomena involving some quantities with dimensions and we assume that a system of 
dimensional units is given throughout. If we are studying physical phenomena, 
we will need the SI system or the MKS system, for example.
In this paper we denote the dimension of a quantity (variable) $\phi$ by $\mathbb{D}(\phi)$ 
instead of the conventional notation $[\phi]$, 
since we are concerned only with relations between dimensions of quantities in general 
rather than specific physical laws. \\
\indent
Let $\{A, B, \ldots\,\!, C\}$ be a minimal system of dimensional units for describing physical phenomena 
with which we are concerned. This means that any dimensional quantity $\phi$
has a dimension represented by $\mathbb{D}(\phi) = A^aB^b \cdots C^c$, 
where $a, b, \ldots\,\!, c$ are all rational numbers,
which are uniquely determined by the dimension of $\phi$. 
This assumption is wide enough for dimensional analysis to be applied elsewhere. 
The following six properties are trivial in physics and we take it for granted that they are valid 
not only in physics but in other disciplines that deal with dimensional quantities. 
These provide us with our starting point of the discussions and will be freely utilized. 
\begin{itemize}
\item[(i)] For two quantities $\phi$ and $\psi$, the addition $\phi+\psi$ or the subtraction $\phi-\psi$
can be defined adequately, if and only if 
$\mathbb{D}(\phi)=\mathbb{D}(\psi)$ holds, i.e., $\phi$ and $\psi$ have the same dimension.
\item[(ii)] If $\mathbb{D}(\phi) = A^aB^b \cdots C^c$ and $k$ is a rational number, 
then $\mathbb{D}(\phi^k) = A^{ka}B^{kb} \cdots C^{kc}$.
\item[(iii)] If $\mathbb{D}(\phi) = A^aB^b \cdots C^c$ and $\mathbb{D}(\psi) = A^dB^e \cdots C^f$, then
$\mathbb{D}(\phi \psi) = A^{a+d}B^{b+e} \cdots C^{c+f}$. 
\item[(iv)] For a dimensionless quantity $\phi$, we have $\mathbb{D}(\phi) = A^0B^0 \cdots C^0$. 
We will denote it simply by $\mathbb{D}(\phi)=1$.
Otherwise we call $\phi$ dimensional or a quantity having a dimension, sometimes
denoted by $\mathbb{D}(\phi) \ne 1$.
\item[(v)] For a natural number $n$, if the $n$th derivative $d^n{\psi}/d{\phi}^n$ is defined, then 
$\mathbb{D}(d^n{\psi}/d{\phi}^n) = \mathbb{D}(\psi {\phi^{-n}})$.
\item[(vi)] If $\psi$ is a function of $\phi$, then $\mathbb{D}(\int \psi d\phi) = \mathbb{D}(\psi \phi)$.
\end{itemize}
\indent
When we regard $A^aB^b \cdots C^c$ as an algebraic product, (ii) and (iii) can be rewritten as the expressions 
$\mathbb{D}(\phi^k) = (\mathbb{D}(\phi))^k$ and  
$\mathbb{D}(\phi \psi) = \mathbb{D}(\phi)\mathbb{D}(\psi)$, respectively. \\
\indent
Some constants may have their own dimensions; Planck's constant $h$ and Boltzmann's constant $k$, for example, 
are not dimensionless. It should be remarked that constants play an important role in dimensional analysis, too. 
In particular, Planck's constant with dimension of `action' will be mentioned in Section 3. \\
\indent
Another usage of constants is illustrated by the following example. 
Let us consider a second-order differential equation ${\psi}^{\prime\prime} + \psi = 0$ in view of dimensional analysis.
Here $\psi$ is an unknown function of $\phi$. To this end we need set ${\psi}^{\prime\prime} + a\psi = 0$ 
by introducing a constant $a$.
Then, putting $\mathbb{D}(\phi) = A$ and $\mathbb{D}(\psi) = B$, 
we postulate $a=1$ and $\mathbb{D}(a) = A^{-2}$. 
We keep $a$ as such a constant in the course of solving the equation and finally put $a=1$. \\
\indent
The next section is devoted to the investigations of the dimensions for functions 
that yield physical quantities and the dimensions for the arguments 
of the functions. In particular, giving simple examples, we show that there are three types of functions 
from a viewpoint of dimensional analysis. As an illustrative example of application 
we give simple second-order ordinary differential equations. 
We refer to an easily available reference [3] that deals with solutions of first-order
ordinary differential equations by dimensional analysis. 
Our results may be applied to some examples of differential equations treated therein.
The final section supplies some related remarks. \\
\indent
It is easily recognized that the theory proposed in this paper can also be applied to similar circumstances in
other disciplines, {\it e.g.}, engineering, economics or biology, which involve quantities with dimensions. \\
\\
\\
\noindent
 {\bf 2.  Dimensional Analysis and Functions}\\
\\
\indent
In order to describe physical laws we need fundamental functions; 
exponential, logarithmic and trigonometric functions {\it etc}. They are building blocks for
more complicated functions. We examine these functions from a viewpoint of dimensional analysis. \\
\indent
We can classify such functions $\psi = \psi(\phi)$ into the following three types in view of dimensions of both 
$\phi$ and $\psi$. Note that the first two types of functions can take dimensional quantities as their arguments. 
The examples taken here are the simplest ones exhibiting respective features. 
Even these simple examples have interesting implications. In the next section
a procedure for obtaining functions of respective types will be given that is based on the consideration
of this section. \\
\\
{\bf Type I}. Both $\phi$ and $\psi$ can have dimensions.\\
\indent
A typical and obvious example is the power function $\psi = \phi^r$ with $r$ a non-zero rational number;
$\phi^3$, $\phi^{-1}$ and $\sqrt{\phi}$ {\it etc}. It is obvious that it has a dimension
if the argument $\phi$ has a dimension; $\mathbb{D}(\psi) = (\mathbb{D}(\phi))^r \ne 1$. \\
\\
{\bf Type II}. $\psi$ is dimensionless, whatever the dimension of $\phi$ is.\\
\indent
An important example is the logarithmic function. As for the definition of (natural) logarithmic function we take 
\begin{eqnarray*}
  \psi = \ln \phi = \int_1^{\phi} \frac {d\phi }\phi.
\end{eqnarray*}
Therefore, we see that $\psi$ is dimensionless, since so is the part $\phi^{-1}d{\phi}$, 
whatever the dimension of $\phi$ is. Hence we always have 
\begin{eqnarray}
  \mathbb{D}(\ln \phi) = 1,
\end{eqnarray}
whether $\phi$ has a dimension or not. \\
\indent
There are many laws involving the logarithmic function $\ln \phi$ with $\mathbb{D}(\phi) \ne 1$ in physics, where $\phi$ may have
a dimension of volume or temperature {\it etc}. Using (1), it is easy to see that
functions of the form $\psi = \phi^r \ln \phi$ 
with $r$ non-zero rational numbers are of Type I. \\ 
\\
{\bf Type III}. Both $\phi$ and $\psi$ must be dimensionless; $\mathbb{D}(\phi) = \mathbb{D}(\psi) =1$.\\
\indent
Examples for this type include all of trigonometric functions. We show this by $\psi = \sin \phi$. 
We begin with the integral
\begin{eqnarray}
  \phi = \int_0^{\psi} \frac {d\psi }{\sqrt{a^2- {\psi}^2}} = \int_0^{\psi} \frac {d(\psi/a)}{\sqrt{1- {(\psi/a)}^2}} 
  = \int_0^{\psi /a} \frac {dx}{\sqrt{1- x^2}} = \arcsin {\frac {\psi}a}
  \end{eqnarray}
for a positive constant $a$. In order for $a^2- {\psi}^2$ to be meaningful, $\mathbb{D}(\psi) = \mathbb{D}(a)$ 
must hold and we get
$\mathbb{D}(\psi/a) = 1$, from which we see that $\phi$ must be dimensionless. 
Writing it as $\sin \phi = \psi/a$, it also follows that  
$\mathbb{D}(\sin \phi) = 1$. It is easy to see that all other trigonometric functions are of this type.\\
\indent
Another important example of Type III is the exponential function. 
We have, for some positive constant $a$,  
\begin{eqnarray*}
  \phi = \ln \psi - \ln a = \int_a^{\psi} \frac {d\psi }\psi.
\end{eqnarray*}
Here $\phi$ is dimensionless, as is shown in Type II, and $\mathbb{D}(\psi) = \mathbb{D}(a)$. 
Rewriting this as $\psi /a = e^{\phi}$, we see that $\mathbb{D}(e^{\phi}) = 1$. \\
\indent
\\
\indent
From the discussion of Type III we can observe that whenever the exponential function $\exp\,(\cdot)$ 
appears in physical laws, the argument is always dimensionless. In statistical mechanics a term
$\exp(-E/kT)$ plays a central role, where both $E$ and $kT$ have a dimension of energy.
So the argument $-E/kT$ is dimensionless. Similarly, wherever $\sin\,(\cdot)$ and $\cos\,(\cdot)$ {\it etc.}
appear in physical laws, their arguments must also be dimensionless. Another illustrative 
example is given by the following. \\
\\
{\bf Example}. Let a quantity $\psi$ be a function of a quantity $\phi$ and let it be
governed by a law 
\begin{eqnarray*}
{\psi}^{\prime\prime} - a^2\psi = 0, 
\end{eqnarray*}
where $a$ is a non-zero constant. 
It is sufficient to introduce two dimensional units $\{A, B\}$; $\mathbb{D}(\phi) = A$ and $\mathbb{D}(\psi) = B$.
Then we get $\mathbb{D}(a) = A^{-1}$ and $a\phi$ is the simplest dimensionless quantity.
The solution is of the form $\psi(\phi) = \alpha \exp\, (a\phi) + \beta \exp\, (- a\phi)$ 
for some constants $\alpha$ and $\beta$,
where $\mathbb{D}(\alpha) = \mathbb{D}(\beta) = B$. \\
\indent
On the contrary the solution of the equation
\begin{eqnarray*}
{\psi}^{\prime\prime} + a^2\psi = 0
\end{eqnarray*}
is of the form $\psi(\phi) = \alpha \sin a\phi + \beta \cos a\phi$, 
where $\alpha$ and $\beta$ are the same as above. \\
\\
\\
\noindent
{\bf 3.  Some Remarks}\\
\\
\indent
From the viewpoint of dimensional analysis Planck's constant $\hslash$ plays an intrigue role 
in producing several dimensional operators in combination with differential operators that are called `observables' in 
quantum mechanics. 
Denoting the dimensions of length, time and mass by $L$, $T$ and $M$, respectively, its 
dimension is that of `action'; $\mathbb{D}(\hslash) = ML^2T^{-1}$. 
Let $(x, y, z )$ be the space variables and $t$ the time variable. Then by the rules of Section 1, we see that
$\mathbb{D}(\hslash \partial/{\partial x})$,
$\mathbb{D}(\hslash \partial/{\partial y})$ and $\mathbb{D}(\hslash \partial/{\partial z})$ 
are all identical to that of the dimension of `linear momentum' $(p_x, p_y, p_z )$ of classical mechanics
and $\mathbb{D}(\hslash \partial/{\partial t})$ is identical to that of the dimension of `energy' $E$.
Thus the correspondence principle of quantum mechanics tells us that classical variables and the corresponding
operators (precisely, $\hslash$ times differential operators) also have the same dimensions as follows; 
\[ p_x \longrightarrow \hslash \partial/{\partial x}, ~~p_y \longrightarrow \hslash \partial/{\partial y},
~~p_z \longrightarrow \hslash \partial/{\partial z}, ~~E \longrightarrow \hslash \partial/{\partial t}.    
\]
On the other hand the correct correspondence should be
\[ p_x \longrightarrow -i\hslash \partial/{\partial x}, ~~p_y \longrightarrow -i\hslash \partial/{\partial y},
~~p_z \longrightarrow -i\hslash \partial/{\partial z}, ~~E \longrightarrow i\hslash \partial/{\partial t}.    
\]
The point is where the imaginary unit $i$ comes from.  \\
\indent
Next we consider a procedure for obtaining functions of Type II or III is considered
by means of the results in the previous section, and finally a remark will be made on the relation to the 
$\Pi$-theorem in dimensional analysis. It is not a general procedure for the derivation of these functions. \\
\indent
Recalling the fact that the logarithmic function $\ln\,(\cdot)$ can take a dimensional quantity as the argument, 
we are able to derive some functions of Type II. Let us consider the function, for instance,
\begin{eqnarray*}
    \psi =  \int_1^{\phi} \frac {\sin (\ln \phi) d\phi }{\phi}.
\end{eqnarray*}
By (1) and the fact that $\sin\,(\cdot)$ is dimensionless, 
we see that $\sin (\ln \phi)$ becomes dimensionless. Hence, using a similar reasoning employed in Type II, we 
get $\mathbb{D}(\psi) = 1$ for this function $\psi$, whether $\phi$ has a dimension or not. \\
\indent
In order to present other functions of Type III, let us define the following, for a natural number $n$, 
\begin{eqnarray*}
  y = \int_0^x \frac {dx}{\sqrt[n]{1- x^n}}  = G_n(x)
\end{eqnarray*}
and its inverse function $y = F_n(x)$. Considering the corresponding equalities in (2) for this case, 
we get the formula
\begin{eqnarray*}
  \phi = \int_0^{\psi} \frac {d\psi}{\sqrt[n]{a^n- \psi^n}}  = G_n(\frac{\psi}a).
\end{eqnarray*}
In a similar manner utilized for trigonometric functions (Type III) in section 2 we can obtain 
$\mathbb{D}(a) = \mathbb{D}(\psi)$ and $\mathbb{D}(\sqrt[n]{a^n- \psi^n}) = \mathbb{D}(\psi)$.
Hence we get $\mathbb{D}(\phi) = 1$. Thus, from $\psi /a = F_n(\phi)$, we can obtain 
functions $F_n$ of Type III; $\mathbb{D}(\phi) = \mathbb{D}(F_n(\phi))=1$. 
They are generalizations of trigonometric functions. Indeed, when the mathematician C. F. Gauss (1777-1855)
attempted to generalize trigonometric functions, he at first introduced and 
considered these functions $G_n$ and $F_n$ prior to 
the discovery of elliptic functions. \\
\indent
When conventionally dealing with dimensional analysis, we often utilize the $\Pi$-theorem ([1, 2]), 
which requires our efforts to find a function
\[
\psi = \Pi(\phi_1, \phi_2, \ldots\,\!,\phi_m).
\]
Here all quantities together with $\psi$ are dimensionless; $\mathbb{D}(\phi_1) = \cdots = \mathbb{D}(\phi_m) =
\mathbb{D}(\psi)=1$. 
Although it is a function of several variables,
it seems that our results may be helpful to determine its form. \\
\\
\\
\noindent
{\bf References}
\begin{itemize}
\item[{[1]}]    G. I. Barenblatt, "Dimensional Analysis", Gordon and Breach, New York, 1987.
\item[{[2]}]    G. I. Barenblatt, "Scaling, Self-Similarity and Intermediate Asymptotics", 
Cambridge University Press, 1996.
\item[{[3]}]  J. A. Belinch\'on, Ordinary differential equations through dimensional analysis,
arXiv: physics/0502154.
\end{itemize}

\end{document}